\magnification 1200
\nopagenumbers
\input boxedeps.tex 
\input boxedeps.cfg 
\SetDefaultEPSFScale{600}
\baselineskip=7mm
\hsize=5.0in

\centerline {\bf { Conformal Symmetry and Triviality}}
\vskip .2truein
\centerline {\it{Talk given at the Meeting for the 60'th Birthday of t.u. at ITU on Jan. 6th, 2003}}
\vskip .7truein
\centerline { M.Horta\c csu * }
\vskip .4truein
\centerline {  Physics Department, Faculty of Sciences and Letters}
\vskip .1truein
\centerline { ITU 80626 Maslak, Istanbul, Turkey }
\vskip .1truein
\centerline {and}
\vskip .1truein
\centerline { Feza G\" ursey Institute, \c Cengelk\" oy, 81220, Istanbul, Turkey}
\vskip 1.3truein
\centerline {\bf { Abstract}}
We study examples where conformal invariance implies  triviality of the underlying quantum field theory.

\vskip .3truein
\noindent
key words: rational critical indices, triviality, conformal symmetry   
\vskip 1.5truein
\noindent
* hortacsu@itu.edu.tr

\baselineskip=18pt
\footline={\centerline{\folio}}
\pageno=1
\vfill\eject

\noindent
Having  conformal symmetry in a quantum field theoretical model implies that the Lagrangian of the model does not possess any dimensional parameters.  The mass term  or any dimensional coupling constant is sufficient to break this symmetry. Since any cut-off term also means a dimensional parameter, the model should be free from infinities. It  should not be necessary  to renormalize the theory in any fashion.

The fulfillment of these  restrictions still does not guarantee that the model has conformal symmetry. Unless the space-time dimensions equals two, we have to make sure that the two point Wightman function of the model should be  $^{/1}$ an integer power of  $$(x-y)^2+i\epsilon(x_{0}-y_{0}). $$ In two dimensions, this function is given as logarithm of   $(x-y)^2+i\epsilon(x_{0}-y_{0}) $ for scalar fields.  If we use a non-integer power of this term $^{/2}$ the index comes out of the log function as a coefficient, still keeping  the same form.  In higher dimensions a power of the same function   have to be used.  If this power is not an integer, we can change the sign between the two terms in $(x-y)^2+i\epsilon(x_{0}-y_{0})$ by a conformal transformation. If the power of this expression is an integer, we can seperate this term into its real ( Principal value)  and  imaginary (delta function) parts and show that under conformal transformations both terms transform covariantly, keeping the sign as it was before the transformation.  If the power is not an integer, this seperation is not possible. In all these remarks we wanted to maintain conformal symmetry in the Wightman function sense.

It was shown $^{/3}$ that if this power is a rational number, you can carry the local conformal symmetry to the global one in the covering space in the general sense.  If the power is an irrational number, even this is not possible by using a countable number of sheets in the covering space.

In $1+1$ dimensions spin  is a parameter which can take any value.  The presence of conformal symmetry was shown to be present in Thirring model for discrete values of spin and coupling constant $^{/4}$.  Similar behaviour is also seen in the analysis of the conformal group in $1+1$ dimensions if one uses "analytical representations" $^{/5}$  .

The rational indices in physics also occur in statistical mechanical models  when conformal
symmetry is imposed. It is found that $^{/6}$  the indices used in phase transitions should be discrete and rational if we want to satisfy the unitarity constraint. One can not extend this construction to relativistic quantum field theoretical models, though, and show the existence of non-trivial models in this field.  The question is whether the rationality of the different powers (indices) that describes these models also makes these models trivial.  Much work was done when the central charge $c$ is equal to unity. I do not recall work when $c>1$ in relativistic quantum field theoretical models using local fields, but this may be only due to  my ignorance on this field.

Schroer $^{/7}$ remarks that if a model is built in Minkowski space, imposing conformal symmetry on the model may result in making the model "trivial". 
He also points out to the fact that in anti-de Sitter (AdS) spaces, however, the Hamiltonian may be replaced by a generator of the symmmetry, which may make a non-trivial construction possible. He also remarks that this construction seems improbable $^{{8}}$.
 The connection between AdS spaces and conformal symmetry was shown in the early nineties $^{/9}$.  Maldecena made the subject popular a little later $^{/10}$. As far as I know it is an open problem whether one can construct non-trivial generic models   higher than two Lorentzian dimensions with global conformal symmetry. We all  know that the only example in $d=4$, $ N= 4 $ super Yang-Mills theory  exists.
We want to find other examples.

By "triviality" we mean models where interactions vanish as the cut-off is removed. A famous example is the Baker-Kincaid paper on the  $\phi^4$ model  in four dimensions $^{/11}$.   Ken Wilson pointed out to  similar phenomena in $d=4$ earlier $^{/12}$. Klauder's recent work shows that this problem  is not completely solved yet $^{/13}$ .

Additional references can be found in the thesis of Reenders $^{/14}$,  submitted to University of Groningen.  Here it is shown how the Nambu-Jona-Lasinio model $^{/15}$  goes  to a  trivial model , since the coupling constant  is proportional to a negative power of the cut-off or the correlation length.  It is essential that the space-time dimension is four or larger for this to occur. For $d=3+1$ , the coupling constant is proportional to a negative power of the logarithm of the cut-off. For higher dimensions, it is proportional to a negative power of the cut-off.
As the cut-off or the correlation length goes to infinity, the interaction vanishes.

A related phenomenon was shown in quantum electrodynamics (QED) long ago $^{/16}$.  Here one shows how the virtual fermion-antifermion pairs form dipoles which screen the charge.

Gell-Mann and Low $^{/17}$ stated that to have a non-trivial theory, the derivative of the coupling constant with respect to the cutoff, $$ \beta = \Lambda {{dg}\over{d\Lambda}} $$ should have a non-trivial fixed point. The two $^{/18}$ and approximate three-loop $^{/19}$ calculations of QED were consistent with the triviality of the model. Attempts to go beyond perturbation theory $^{/20}$
did not bring any results that contradict this fact. The fact that QED results are confirmed by experiments in an excellent manner makes many forget that QED  is actually an asymptotic model.

Additional work on his problem was done by many groups.  This work is described in Reenders' thesis $^{/14}$.  Bardeen and co-workers $^{21}$ showed that in the " quenched ladder" approximation of QED , the dimension of the four fermion operator is no longer six but four at a critical coupling.  This fact makes it necessary to add Nambu-Jona-Lasinio type terms to the QED Lagrangian, since these terms have the same dimensions as the interaction of the model. Then the new model , named as gauged Nambu-Jona-Lasinio model , should replace QED for this range of values of the fermion-photon coupling  constant. Groups led by Miransky $^{/22}$, Appelquist $^{/23}$ and Kogut $^{/24}$  further studied this model and confirmed these results.  Lattice calculations $^{/24}$  also confirm that pure QED is trivial.  Reenders studies the gauged Nambu-Jona-Lasinio model with internal symmetry using non-perturbative methods $^{/26}$, and concludes that if there are more than 44 ( in another approximation 54 ) flavors of fermions,  scalar and pseudo-scalar bound-states made out of fermions will prevent the total screening of the charge, resulting in a non-trivial conformal invariant model. 

Similar examples of "triviality"  were studied in the eighties in fermionic models with some form of conformal symmetry $^{/27}$ These models were quantized using ${{1}\over{N}}$ methods resulting in effective lagrangians as 
$$ L_{eff} = \overline{\psi} \gamma^{\mu} ( \partial_{\mu} -m -e A_{\mu}) \psi . $$ 
The bosonic kinetic term wis generated by the one-loop correction, which also shows that the coupling constant goes to zero as the cut-off is removed.
The question was whether inifinities in higher order calculations will cancel this zero, resulting in finite results for physical processes.  Two and three loop calculations for several processes showed $^{/28}$ that indeed a finite result was possible, but it was exactly equal to the naive- quark result.  One could not obtain the logarithmic corrections.

One may deduce that imposing conformal symmetry makes these non-linear models "trivial".  One may still add additional terms to these models, study the behaviour of their beta functions, and see if there are non-trivial fixed points of these models.  I think this will be a worthwhile endeavor to follow up.
\vskip .1truein
\noindent
{\bf{Acknowledgement}}
I thank Prof.Dr. A.Nihat Berker, the chairman of the physics department at ITU and Dr. S.Kayhan  \" Ulker for organizing this meeting.  

\vskip .3truein
\noindent
{\bf{References}}
\item {1.}M.Horta\c csu, R. Seiler and B.Schroer, Phys. Rev.D {\bf{5}}, (1972)2519;

\item {2.} K.Higashijima and E.Itou, hep-th/0303090 (2003) ;

\item {3.} B.Schroer and J.A.Swieca, Phys. Rev. D {\bf{10}} (1974) 480, B.Schroer, J.A.Swieca and A.H.Voelkel , Phys. Rev. D {\bf{11}} (1975) 11;

\item {4.} M.Horta\c csu, Il Nuovo Cimento {\bf{17A}} (1973) 411;

\item {5.}  G.G\" ursey and S.Orfanidis, Phys. Rev.D {\bf{7}} (1973) 2414;

\item {6.} D.Friedan, Z. Qui and S.Shenker, Phys. Rev. Lett. {\bf{52}} (1984) 1575 ;

\item {7.}  B.Schroer, Phys. Lett. B {\bf{506}} (2001) 337;

\item {8}  B.Schroer, J. Phys. A: Math. Gen {\bf{34}} (2001) 3638;

\item {9.}  G.T.Horowitz and D.L.Welch, Phys. Rev. Lett. {\bf{71}} (1993) 328; N.Kaloper, Phys. Rev. D {\bf{48}} (1993) 2598;

\item {10.}  J.Maldecena, Adv. Theo. Math. Phys. {\bf{2}} (1998) 231;

\item {11}  G.A.Baker and J.M. Kincaid, Phys. Rev. Lett. {\bf { 42}} (1979) 1431; J. Statistical Phys. {\bf{24}} (1981) 469 ;

\item {12} K.G. Wilson, Phys. Rev. D {\bf{7}} (1973) 2911;

\item {13}  J.Klauder, hep-th/0209177, v3-Dec.12, 2002;

\item {14} M.Reenders, hep-th/9906034;

\item {15} Y.Nambu and G.Jona-Lasinio, Phys. Rev. {\bf{122}} (1961) 345;

\item {16}  L.D.Landau and I.Ya. Pomeranchuk, Doklady Akad. Nauk, {\bf{102}} (1955) 489;

\item {17}  M. Gell-Mann and F.E.Low, Phys. Rev. {\bf{95}} (1954) 1300;

\item {18}  R.Jost and J.M.Luttinger, Helv. Phys. Acta, {\bf{23}} (1950) 201;

\item {19} J.L.Rosner, Phys. Rev. Lett. {\bf{17}} (1966) 1190;

\item {20}  K.Johnson, R. Willey and M. Baker, Phys. Rev. {\bf{163}} (1967) 1699; M.Baker and K.Johnson, Phys. Rev. {\bf{183}} (1969)1292, Phys. Rev. D {\bf{8}} (1973) 1110; S.L. Adler , Phys. Rev. D {\bf{5}} (1972) 3021; S.L.Adler, C.G.Callan, D.J.Gross and R. Jackiw, Phys. Rev. D {\bf{6}} (1972) 2982; M. Baker and K.Johnson, Physica A {\bf{96}} (1979) 120; 

\item {21} W.A.Bardeen, C.N.leung and S.T.Love, Phys. Rev. Lett. {\bf{56}} (1986) 1230, Nuclear Physics B {\bf{ 273}} (1986) 649;

\item {22}  V.A.Miransky and K.Yamawaki, Mod. Phys. Lett. A {\bf{4}} (1989) 129,  Phys. Rev. D {\bf{55}} (1997) 5051, additional information can be obtained from, V.A.Miransky, {\it{Dynamical Symmetry Breaking in Quantum Field Theories}}, World Scientific, Singapore (1993);

\item {23} T.Appelquist, J.Terning and L.C.R. Wijewardhana, Phys. Rev D {\bf{44}} (1991) 871;

\item {24}  A.Kocic, S.Hands, J.B.Kogut and E.Dagotto, Nucl. Phys. B {\bf{347}} (1990) 217; A.Kocic, J.B.Kogut, M.P. Lombardo and K.C.Wang, Nucl. Phys.B {\bf{ 397}} (1993) 409;

\item {25}  M.Gockeler, R.Horsley, V. Linke, P.E.L.Rakow, G.Schierholz and H.Stuben, Phys. Rev. Lett. {\bf{80}} (1998) 4119;

\item {26} M.Reenders, Phys. Rev.D {\bf{62}} (20009 025001;

\item {27}  K.G. Akdeniz, M. Ar\i k, M.Durgut, M.Horta\c csu, S.Kaptano\u glu, N.K.Pak, Phys. Lett. B {\bf{116}} (1982) 34, 41, K.G. Akdeniz, M. Ar\i k, M.Horta\c csu, N.K.Pak, Phys. Lett. B {\bf{124}} (1983) 79;

\item {28}   M. Ar\i k and M.Horta\c csu, J. Phys. G : Nuclear Phys. {\bf{9}} (1983) L119.
\end